\documentclass[twoside,twocolumn,9pt]{article}
\usepackage{extsizes}
\usepackage[super,sort&compress,comma]{natbib} 
\usepackage[version=3]{mhchem}
\usepackage[left=1.5cm, right=1.5cm, top=1.785cm, bottom=2.0cm]{geometry}
\usepackage{balance}
\usepackage{times,mathptmx}
\usepackage{sectsty}
\usepackage{graphicx} 
\usepackage{lastpage}
\usepackage[format=plain,justification=justified,singlelinecheck=false,font={stretch=1.125,small,sf},labelfont=bf,labelsep=space]{caption}
\usepackage{float}
\usepackage{fancyhdr}
\usepackage{fnpos}
\usepackage[english]{babel}
\usepackage{array}
\usepackage{droidsans}
\usepackage{charter}
\usepackage[T1]{fontenc}
\usepackage[usenames,dvipsnames]{xcolor}
\usepackage{setspace}
\usepackage[compact]{titlesec}

\usepackage{amsfonts}
\usepackage{bm}

\usepackage{epstopdf}%This line makes .eps figures into .pdf - please comment out if not required.

\definecolor{cream}{RGB}{222,217,201}

\begin{document}

\pagestyle{fancy}
\thispagestyle{plain}
\fancypagestyle{plain}{

%%%HEADER%%%
%\fancyhead[C]{\includegraphics[width=18.5cm]{head_foot/header_bar}}
%\fancyhead[L]{\hspace{0cm}\vspace{1.5cm}\includegraphics[height=30pt]{head_foot/PCCP}}
%\fancyhead[R]{\hspace{0cm}\vspace{1.7cm}\includegraphics[height=55pt]{head_foot/RSC_LOGO_CMYK}}
\renewcommand{\headrulewidth}{0pt}
}
%%%END OF HEADER%%%

%%%PAGE SETUP - Please do not change any commands within this section%%%
\makeFNbottom
\makeatletter
\renewcommand\LARGE{\@setfontsize\LARGE{15pt}{17}}
\renewcommand\Large{\@setfontsize\Large{12pt}{14}}
\renewcommand\large{\@setfontsize\large{10pt}{12}}
\renewcommand\footnotesize{\@setfontsize\footnotesize{7pt}{10}}
\makeatother

\renewcommand{\thefootnote}{\fnsymbol{footnote}}
\renewcommand\footnoterule{\vspace*{1pt}% 
\color{cream}\hrule width 3.5in height 0.4pt \color{black}\vspace*{5pt}} 
\setcounter{secnumdepth}{5}

\makeatletter 
\renewcommand\@biblabel[1]{#1}            
\renewcommand\@makefntext[1]% 
{\noindent\makebox[0pt][r]{\@thefnmark\,}#1}
\makeatother 
\renewcommand{\figurename}{\small{Fig.}~}
\sectionfont{\sffamily\Large}
\subsectionfont{\normalsize}
\subsubsectionfont{\bf}
\setstretch{1.125} %In particular, please do not alter this line.
\setlength{\skip\footins}{0.8cm}
\setlength{\footnotesep}{0.25cm}
\setlength{\jot}{10pt}
\titlespacing*{\section}{0pt}{4pt}{4pt}
\titlespacing*{\subsection}{0pt}{15pt}{1pt}
%%%END OF PAGE SETUP%%%

%%%FOOTER%%%
\fancyfoot{}
%\fancyfoot[LO,RE]{\vspace{-7.1pt}\includegraphics[height=9pt]{head_foot/LF}}
%\fancyfoot[CO]{\vspace{-7.1pt}\hspace{11.9cm}\includegraphics{head_foot/RF}}
%\fancyfoot[CE]{\vspace{-7.2pt}\hspace{-13.2cm}\includegraphics{head_foot/RF}}
%\fancyfoot[RO]{\footnotesize{\sffamily{1--\pageref{LastPage} ~\textbar  \hspace{2pt}\thepage}}}
%\fancyfoot[LE]{\footnotesize{\sffamily{\thepage~\textbar\hspace{4.65cm} 1--\pageref{LastPage}}}}
\fancyhead{}
\renewcommand{\headrulewidth}{0pt} 
\renewcommand{\footrulewidth}{0pt}
\setlength{\arrayrulewidth}{1pt}
\setlength{\columnsep}{6.5mm}
\setlength\bibsep{1pt}
%%%END OF FOOTER%%%

%%%FIGURE SETUP - please do not change any commands within this section%%%
\makeatletter 
\newlength{\figrulesep} 
\setlength{\figrulesep}{0.5\textfloatsep} 

\newcommand{\topfigrule}{\vspace*{-1pt}% 
\noindent{\color{cream}\rule[-\figrulesep]{\columnwidth}{1.5pt}} }

\newcommand{\botfigrule}{\vspace*{-2pt}% 
\noindent{\color{cream}\rule[\figrulesep]{\columnwidth}{1.5pt}} }

\newcommand{\dblfigrule}{\vspace*{-1pt}% 
\noindent{\color{cream}\rule[-\figrulesep]{\textwidth}{1.5pt}} }

\newcommand*{\citen}[1]{%
  \begingroup
    \romannumeral-`\x % remove space at the beginning of \setcitestyle
    \setcitestyle{numbers}%
    \cite{#1}%
  \endgroup   
}

\makeatother
%%%END OF FIGURE SETUP%%%

%%%TITLE, AUTHORS AND ABSTRACT%%%
\twocolumn[
  \begin{@twocolumnfalse}
\vspace{3cm}
\sffamily
\begin{tabular}{m{1.5cm} p{13.5cm} }

& \noindent\LARGE{\textbf{Predicting outcomes of catalytic reactions using machine learning$^\dag$}} \\%Article title goes here instead of the text "This is the title"
\vspace{0.3cm} & \vspace{0.3cm} \\

 & 
 \noindent\large{Trevor David Rhone,$^{\ast}$\textit{$^{a}$} 
 Robert Hoyt,\textit{$^{a}$}  
 Christopher R. O'Connor,\textit{$^{b}$}
 Matthew M. Montemore,\textit{$^{b,c}$} 
 Challa S.S.R. Kumar,\textit{$^{b,c}$}
 Cynthia M. Friend,\textit{$^{b,c}$}
 and 
 Efthimios Kaxiras \textit{$^{a,c}$}} \\

& 

 \noindent\normalsize{Predicting the outcome of a chemical reaction using efficient computational models can 
  be used to develop high-throughput screening techniques.
  %which can save significant time and effort over performing a large body of experiments. 
  This can significantly reduce the number of experiments needed to be performed in a huge search space, which saves time, effort and expense.
  Recently, machine learning methods have been bolstering conventional structure-activity relationships used to advance understanding of chemical reactions.
  We have developed a model to predict the products of catalytic reactions on the surface of oxygen-covered and bare gold using machine learning. Using experimental data, we developed a machine learning model that maps reactants to products, using a chemical space representation. 
  This involves predicting a chemical space value for the products, and then matching this value to a molecular structure chosen from a database.
  The database was developed by applying a set of possible reaction outcomes using known reaction mechanisms. Our machine learning approach complements chemical intuition in predicting the outcome of several types of chemical reactions. In some cases, machine learning makes correct predictions where chemical intuition fails.
  We achieve up to 93$\%$ prediction accuracy for a small data set of less than two hundred reactions.} \\%The abstrast goes here instead of the text "The abstract should be..."

\end{tabular}

\end{@twocolumnfalse} \vspace{0.6cm}
]
%%%END OF TITLE, AUTHORS AND ABSTRACT%%%

%%%FONT SETUP - please do not change any commands within this section
\renewcommand*\rmdefault{bch}\normalfont\upshape
\rmfamily
\section*{}
\vspace{-1cm}

%%%FOOTNOTES%%%

\footnotetext{\textit{$^{a}$~Department of Physics, Harvard University, Cambridge, Massachusetts 02138, USA. Fax: 617 495 0416; Tel: 617 495 2872; E-mail: kaxiras@g.harvard.edu}}
\footnotetext{\textit{$^{b}$~Department of Chemistry and Chemical Biology, Harvard University, Cambridge, Massachusetts 02138, USA. }}
\footnotetext{\textit{$^{c}$~School of Engineering and Applied Sciences, Harvard University, Cambridge, Massachusetts 02138, USA }}

%Please use \dag to cite the ESI in the main text of the article.
%If you article does not have ESI please remove the the \dag symbol from the title and the footnotetext below.
\footnotetext{\dag~Electronic Supplementary Information (ESI) available: See DOI: 10.1039/cXCP00000x/}
%additional addresses can be cited as above using the lower-case letters, c, d, e... If all authors are from the same address, no letter is required

%\footnotetext{\ddag~Additional footnotes to the title and authors can be included \textit{e.g.}\ `Present address:' or `These authors contributed equally to this work' as above using the symbols: \ddag, \textsection, and \P. Please place the appropriate symbol next to the author's name and include a \texttt{\textbackslash footnotetext} entry in the the correct place in the list.}

%%%END OF FOOTNOTES%%%

%%%%%%%%%%%%%%%%%%%%%%%%%%%%%%%%%%%%%%%%%%%%%%%%%%%%%%%%%%%%%%%%%%%%%
%% The document title should be given as usual. Some journals require
%% a running title from the author: this should be supplied as an
%% optional argument to \title.
%%%%%%%%%%%%%%%%%%%%%%%%%%%%%%%%%%%%%%%%%%%%%%%%%%%%%%%%%%%%%%%%%%%%%
% \title[Predicting outcomes of catalytic reactions using machine learning]
%   {}%\footnote{A footnote for the title}}

% \abbreviations{Au, ML}
% \keywords{Au catalysis, machine learning}

% \begin{document}

\section{Introduction}

\begin{figure*}[ht]
\begin{center}
  \includegraphics[width=0.78\textwidth, keepaspectratio]{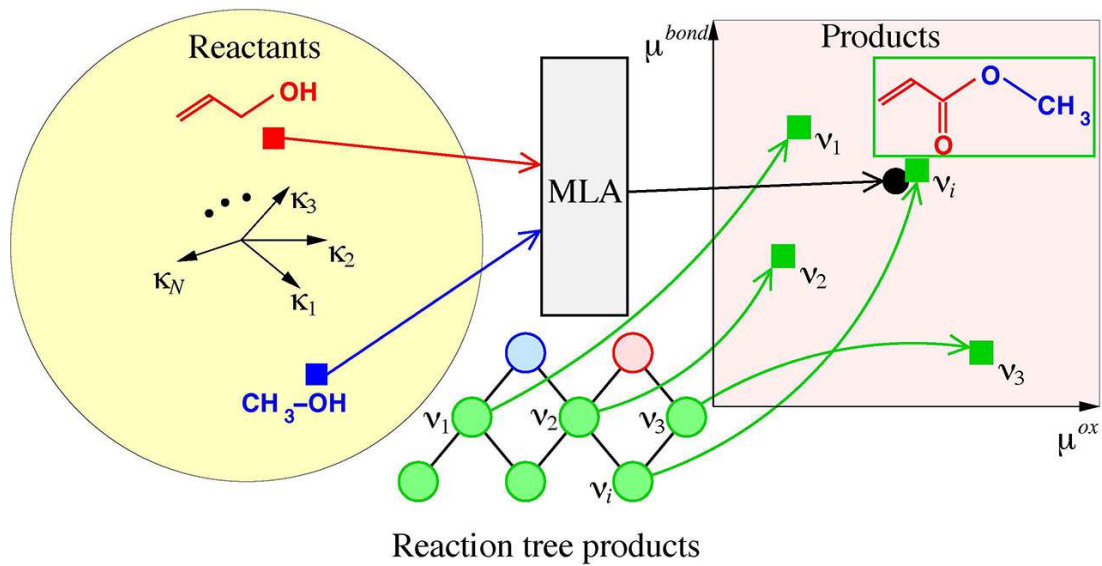}
%   \caption{ \label{fig1} Schematic of the machine learning process used to predict the outcome of heterogeneous
%   catalysis. Reactants are shown in blue and red.
%   Orange markers represent the reaction outcome.
%   Squares (circles) represent chemical space values (molecular structures).}
  \caption{\label{fig1} 
  Schematic representation of the ML process used to predict the outcome of catalytic reactions on the oxygen-covered or bare Au surface.  
  Reactants (blue and red squares) from $N$-dimensional reactant space (axes labeled $\kappa_1, \kappa_2, \dots, \kappa_N$)
are mapped through the machine learning algorithm 
(MLA, grey shaded box) to products (black square)
on the two-dimensional product space represented by
($\mu^{ox}, \mu^{bond}$).  
In conjunction with this mapping, a Reaction Tree is used to generate many
molecular structures from the reactants (blue and red circles) to 
possible reaction outcomes (green circles), 
which correspond to specific values on the ($\mu^{ox}, \mu^{bond}$) plane, 
shown as $\nu_1, \nu_2, \dots, \nu_i$ (green squares).  
The most likely outcome of a given reaction is the molecular structure (green square) closest to the prediction of the MLA (black square).
}
\end{center}
\end{figure*}
%
%\begin{itemize}
%\item It's important/useful to predict the products of a reaction
Efficient prediction of reaction products has long been a major goal of the organic chemistry community\cite{tropsha2010, Szymkuc2016, aspuru2016} and the drug discovery process in the pharmaceutical industry~\cite{Jorgensen2009,Jorgensen1813}.
These predictions would enable advances in high-throughput screening or molecular inverse design.
While most previous work has focused on solution-phase or gas-phase chemistry, a scheme for the prediction of reaction outcomes would also be useful for heterogeneous catalysis, which can greatly improve the efficiency of a reaction. 
If reasonable predictions can be made for the products that a catalyst is likely to produce from a given set of reactants, synthetic routes can be developed computationally and confirmed experimentally. 
The complex kinetic pathways involving a catalyst make such predictions difficult. 
Moreover, the variability of catalyst conformations, including the
great variety in surface structure, composition, and pressure-dependence adds complexity to this problem.

%\item "Traditional approach"
Traditionally, reaction prediction has relied on chemical intuition. 
While some computational approaches have been developed to aid in reaction prediction, they have generally involved encoding this intuition into a set of rules, and have not been widely adopted\cite{Warr2014,Szymkuc2016,Todd2005,PENSAK1977,Corey1969}. 
This is because it is difficult to enumerate all the chemical rules that exist and perhaps impossible to always place them in the correct chemical context.
It is desirable to develop an approach that does not rely on such rules, to serve as a complement to chemical intuition. 
This type of approach can be particularly useful in cases where the reaction outcome is counterintuitive.

%\item Machine learning can be useful for making predictions without needing a strong understanding of the underlying process.
%
Machine learning (ML) provides an approach to efficient reaction prediction that does not require a strong a priori understanding of the pathways and kinetics involved in the reaction.
Furthermore, inspecting ML models can in some cases aid in the understanding of reaction outcomes.
For instance, in the case of the interaction of oxiranes on bare Au, one may expect that molecular desorption, as is the case for most organics on bare gold.
Instead, what is observed is deoxygenation of oxiranes on bare Au (e.g. 2-methyloxirane $\rightarrow$ prop-1-ene on O/Au(111)).
%That is, machine learning may be able to make accurate predictions in cases that defy chemical intuition. 
In addition, ML approaches have been shown to be more accurate than rule-based approaches on large, general databases~\cite{Segler2017}. While data-driven approaches have been applied to catalysis,~\cite{Okamoto2004,Rothenberg2013,tsuda2016,Liu2017,Hongliang2017,Coley2018,Grajciar2018,Goldsmith2018} previous work has focused on optimizing a catalyst for a particular reaction, not on predicting reaction outcomes.

%\item Drawbacks of previous approaches
ML approaches have been useful in making predictions for organic synthesis,\cite{Engkvist2018,Szymkuc2016,Schneider2015,Warr2014,Marcou2015,Kayala2012a} but they generally require large data sets, and have not been tested in reaction prediction for heterogeneous catalysis. In many cases, millions of reactions are extracted from a chemical database or the United States Patent Office~\cite{Bradshaw2018,Segler2018,Coley2018,Lowe2012}. Even when more limited classes of reactions are studied, data sets with hundreds of thousands of data points are often used~\cite{Carrera2009,Wei2016a,Zhang2005a}. Large data sets with many reactant combinations are available for organic chemistry, but such data sets do not exist for heterogeneous catalysts. Therefore, a new approach is needed for this case.
%that can accurately predict reaction outcomes on a heterogeneous catalyst. To tackle this problem, 
Here we combine machine learning with reaction trees, generated using possible reaction mechanisms, to develop a methodology that does not require very large datasets.

\begin{figure*}[ht ]
  \includegraphics[width=0.90\textwidth, keepaspectratio]{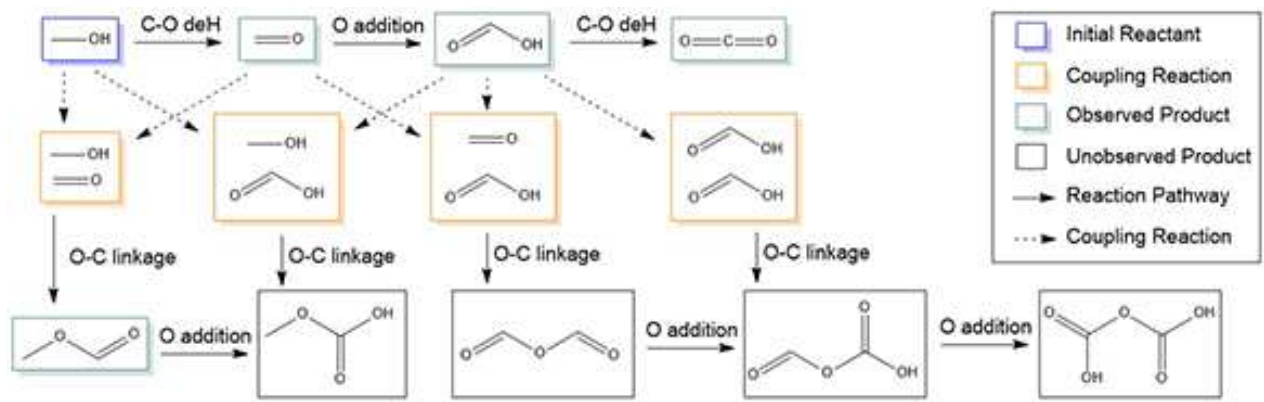}
  \caption{ \label{fig3} Schematic of a reaction tree for methanol. 
  Beginning with methanol as the reactant,
  monomer (coupling) reactions transform a single molecule (pair of molecules) to chemically reasonable products.
  Both the observed product and the unobserved products are shown in the tree.
}
\end{figure*}

%\item Why Au?
Since ML tools are well suited to multivariate analysis of complex relationships, they are appropriate for studying multifunctional catalysts such as Au.
Au is one of the best studied heterogeneous catalysts and is uniquely selective for many reactions, including complex partial oxidative processes. 
Au-based catalysts have been tested for many reactions, often with promising results~\cite{Hughes2005,Personick2016c}. 
However, nanoparticle catalysts or alloy catalysts are quite complicated, and the products formed can depend on a variety of attributes of the material. 
Therefore, we use data gathered on well-defined, single-crystal Au catalysts.
This provides a cleaner data set while still providing the possibility for scaling up, due to the correspondence that is often found between single-crystal Au surfaces and certain nanostructured Au catalysts~\cite{Karakalos2016a}. 

\begin{figure}[ht]
\begin{center}
  \includegraphics[width=0.34\textwidth, keepaspectratio]{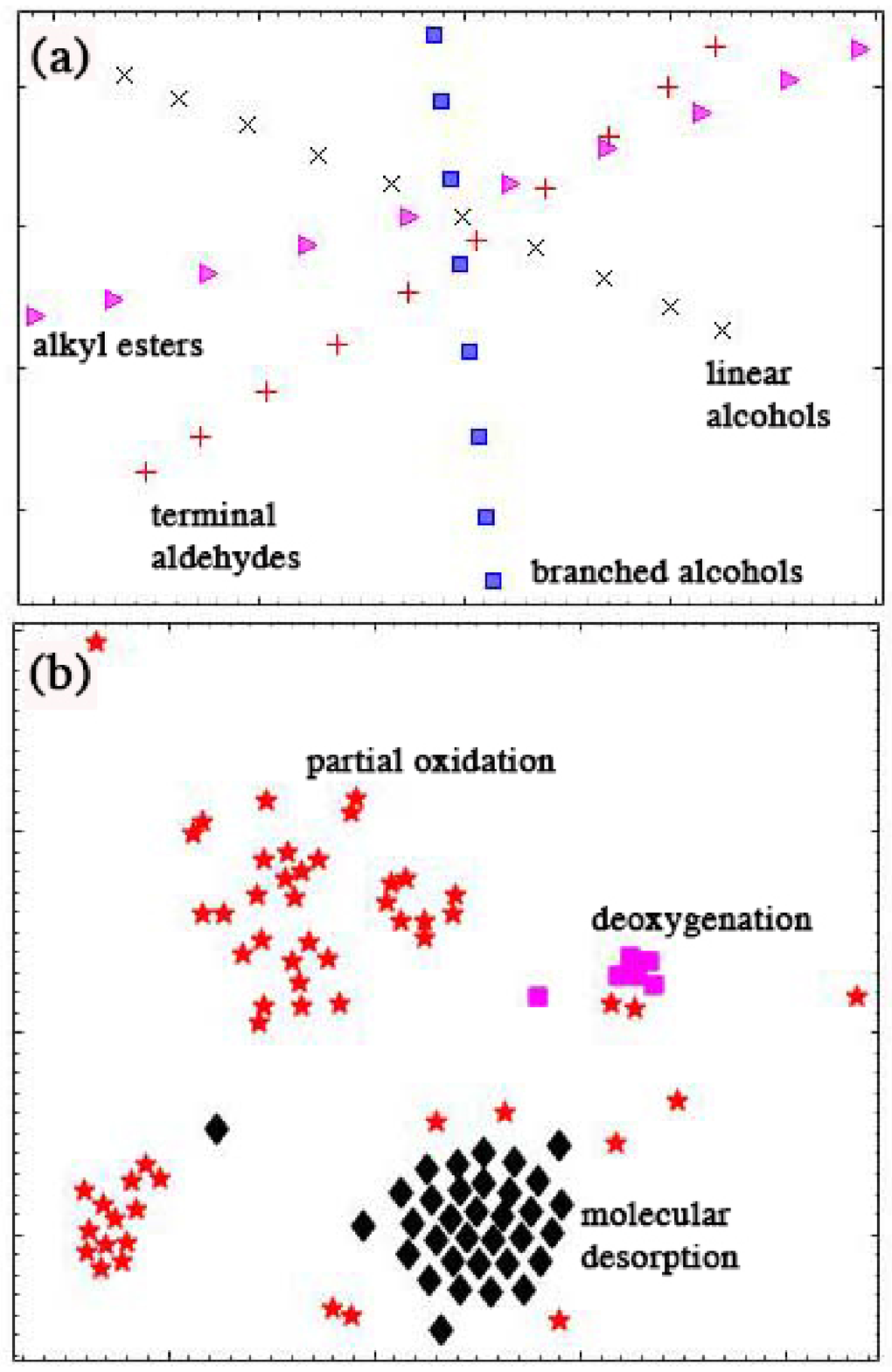}
  \caption{ \label{fig2} 
  Visualization of mathematical representations of molecules and chemical reactions.
  (a) 2D projection of chemical space values (using t-SNE) of several molecular classes of varying carbon chain length: 
  linear alcohols (`x'), 
  branched alcohols (`squares'), 
  terminal aldehydes (`+'), %carboxylic acid (`left triangle'), 
  %methyl esters (`triangle'), 
  alkyl esters (`right triangle').
  The chemical space values based on the bond energy version of the BoB representation is used here.
  (b) 2D projection of $\delta_{rxn}$ 
  using t-SNE. 
  Reaction types are
  %I. complex addition (`o'),
  %II. coupling (`x'), 
  partial oxidation (`star'), 
  %V. hydrogenation (`triangle'), 
  %VI. dehydrogenation (`+'), 
  deoxygenation (`squares') and 
  molecular desorption (`diamond').
  Both the original BoB representation and the BoB representation modified using the oxidation state yield similar results.
  }
\end{center}
\end{figure}

%\item We use ML to make predictions of products for catalytic oxidation on Au surfaces. (Roadmap) How we're different: small dataset, trees, surface reactions; stepping stone to learning routes through trees (chemical insight, including quantitative kinetics).
In this work, we develop a methodology for predicting the outcomes of reactions on Au surfaces, based on an experimental data set with 145 data points. 
We first use ML techniques to predict the products' locations in a chemical space. 
We then use a reaction tree to generate possible products, and select 
the molecule in the tree that is closest to the predicted location in chemical space. 
We achieve up to 93$\%$ prediction accuracy for catalytic reaction outcomes.
In addition to being useful for prediction, this is a first step towards learning the pathways of a reaction, which could be useful for understanding quantitative kinetics and gaining mechanistic insight.
%{

%%%%
\section{Methods}
%%%%
%\begin{itemize}
  %  \item Explain how obtained experimental data - ask Chris to write this portion.
  %  \item Explain how to construct the chemical space values for a set of molecules.
  %  \item Machine learning to predict chemical space values of the products.
  %  \item The inverse problem: Mapping chemical space values to molecular structure
  %  \item Space to expand in the subsections.
%\end{itemize}
%
A schematic of our ML approach to predicting the outcomes of catalytic reactions is shown in Fig.~\ref{fig1}.
Given a set of reactions, we describe the reactants and products of each 
reaction with chemical descriptors
(that is, chemical or atomic properties which describe a molecule). 
In particular, we create a mathematical description of the molecules which contains information about their molecular structure, and use this description to define a chemical space~\cite{tanimoto2015, mbtr,rupp2012,riley2017}.
Molecules are represented as high-dimensional vectors inside this chemical space.
%
%
%We can describe molecules in terms of their chemical similarity to a reference molecule within this space.
%Our definition of molecular representation incorporates both atomic and chemical properties.
%
%We create a molecular representation vector whose components encode information on (i)  atomic oxidation states and (ii) bond energies.
%
We then use ML models to map the molecular representation of the reactants to the chemical space values of the corresponding products.
Predictions are then limited to feasible chemical transformations. 
These transformations of the reactants (i.e. the reaction trees) are generated using our reaction tree method, derived from known reaction mechanisms~\cite{gasteiger1987,aspuru2014}, which are defined below (see the Supplementary Information for details).
The chemical space values of these reaction tree structures are compared with the ML-predicted chemical space values, and the closest molecular structures are selected as the likely reaction products.

\subsection{Experimental data set}
A database of 145
% 184 (old number)
chemical reactions (32 of which are desorption reactions)
on oxygen covered gold (O/Au) and bare gold (Au) was assembled from ultra high vacuum (UHV) studies on Au(111) and Au(110) single crystals (see Supplementary Information for a list of reactions).
\cite{Xu2009,Outka1987,Xu2010,Liu2009,Siler2014,Xu2011,Xu2010a,Xu2012,Deng2005,Deng2006,Xu2013,Xu2010b,Rodriguez-Reyes2014,Xu2011a,Liu2010,Liu2010a,Rodriguez-Reyes2012,Karakalos2016,Zugic2016,Min2009,Outka1987a,Karakalos2016a,Lui,OConnor2018,Jaffey1994,Jaffey1994a,Jaffey1994b,Rodriguez-Reyes2018} The chemical reactions were reported with a reactant and product stoichiometry based on mechanisms determined in the UHV studies. For many reactions, the product distribution is dependent on the initial oxygen coverage. 
To account for this, a collection of a wide range of oxygen coverages (0.05 - 0.5 monolayers) was used to include all reaction products observed experimentally. We did not include reactions which formed radical species because they are not observed under catalytic studies and are only detectable in UHV studies. 
Reactions that lead to complete combustion were also not included, because we chose to focus on predicting value-added chemical transformations. 

The Au(111) and Au(110) surfaces do not dissociate molecular oxygen, so in each experiment strong chemical oxidants (\ce{O3} and \ce{NO_2}) were used to create the oxygen-covered surfaces. 
The chemical reactions were performed by exposure of an organic reactant to a pretreated surface condition at a sufficiently low temperature that no products were immediately desorbed.
The resultant products were determined using temperature programmed reaction spectroscopy (TPRS), where the sample is linearly heated and the desorbed products are detected using a mass spectrometer. 
%The database of chemical reactions is available in the Supplementary Information. 
%%

We have classified the chemical reactions on oxygen-covered gold and bare gold into the following reaction types: I. Complex addition ($\geq$ 3 unique organic reactants), II. coupling (joining of two organic reactants), III. partial oxidation (an increase in the oxidation state of the organic molecule), IV. hydrogenation (addition of hydrogen), V. dehydrogenation (removal of hydrogen), VI. deoxygenation (removal of oxygen) and VII. molecular desorption.

\subsection{Molecular descriptors}

An appropriate representation of the molecules in a chemical reaction is important for quantitatively describing the reaction.
We find that useful chemical descriptors include the atomic oxidation number, the bond energy and the molecular weight
%
%\todo{Did we use all of these? We should list at least a few of the %properties we used.}  --> Yes we included all of them, and also more.
%
(see Supplementary Information for the complete list of chemical properties used to construct the molecular representation).
%Molecular structure can be exploited to design descriptors known as a molecular representations. 
%We used our chemical space to quantify the chemical similarity between molecules and to describe chemical reactions. 
%We use machine learning models to construct rules for mapping reactants to products in chemical %space.
These chemical properties are generally combined with the molecular structure in order to create an overall fingerprint for each molecule. Several molecular representations exist in the literature~\cite{mbtr,rupp2012,riley2017} and have been used to predict quantities like the atomization energy.
They have also been used to predict the outcomes of organic chemistry reactions~\cite{aspuru2016,gambin2017,doyle2018,aspuruguzik2018}. 

We modified the bag of bonds (BoB) representation~\cite{hansen2015} to construct a chemical space that encodes sufficient information to be able to predict the outcome of a catalytic reaction when used alongside ML tools.
Only recently has the BoB descriptor (i.e. the unmodified BoB) been applied
to problems in catalysis~\cite{Terejanu2018}.
The original BoB chemical space distance between two molecules with BoB vectors \textbf{B} and \textbf{B}$_{l}$ is defined as $d(\textbf{B},\textbf{B}_{l})$~\cite{hansen2015}, 
where $d(\textbf{B},\textbf{B}_{l})$ is the distance between vectors represented by the L$_p$-norm,  
($\sum_j | \textbf{B}^j-\textbf{B}_l^j |^p)^{1/p}$.
The entries of the BoB vectors are constructed using the Coulomb kernel~\cite{rupp2012}:
$Z_i Z_j/|R_i - R_j|$,
where $Z_i$ is the nuclear charge and $R_i$ the position of atom $i$, with $i$,$j$ denoting nearest neighbour pairs.

We exploited this framework to define a molecular representation \textbf{M} that includes (i) the oxidation number and (ii) the bond energy instead of the atomic number, as these quantities are more relevant for chemical reactions
\footnote[2]{We obtain the oxidation number from the python package pymatgen~\cite{pymatgen}. The bond energy is the energy to break the bond between a pair of atoms; these values were obtained from tables of bond energies~\cite{nist}.}.
%\todo{Is it literally the bond energy of that bond in that %specific molecule? Where do you get it?}
%
That is, we modified the above expression by substituting the nuclear charge with the oxidation number or the bond energy: 
${\left( |X_i - X_j| \right)/|R_i - R_j|}$, where the numerator can be the difference in the oxidation number or the bond energy of pairs of nearest neighbour atoms.  
We defined the chemical space distance for a molecule as the $p$-norm, d($\textbf{M}$,$\textbf{M}_l$) of the chemical space vector $\textbf{M}$, where $\textbf{M}_l$ is a reference vector containing all zeros. 
We consider the Euclidean distance ($p$=2).
%
%\todo{This originally sounded like the similarity between two molecules is defined as the distance between one of the molecules and the 0 vector. Check to make sure this is now correct.} 
%
We then define a chemical space for the products, $\bm{\mu}=(\mu^{ox},\mu^{bond})$, where $\mu^{ox}$ ( $\mu^{bond}$) is the $p$-norm of the BoB descriptor based on the oxidation number (bond energy).
To represent the reactants, we define an `extended' chemical space vector $\bm{\kappa}$, which is a vector with $\bm{\mu}$ concatenated with other molecular descriptors, such as the $p$-norm of the original BoB representation, the structural complexity~\cite{pubchem}, the molecular weight and the number of surface oxygen atoms
(see Supplementary Information for a complete list of molecular descriptors in the extended chemical space representation).
Molecules that are chemically similar will be generally close in chemical space. 
We used the Pubchem~\cite{pubchem} and pymatgen~\cite{pymatgen} databases to obtain the atomic properties used to construct the chemical space representation.

\subsection{Machine learning methods}
We used ML models to approximate a function which maps the reactants of 
a catalytic reaction to the products, namely, we predicted the chemical space values of the products from a given set of reactants. 
The inputs to our models are the extended chemical space values of the reactants.
%
%\todo{I thought we used a larger feature set for the reactants than just the chemical space values. --> T.R. : Yes that's correct}
%
We considered the following ML methods: lasso, kernel ridge regression and random forest regression~\cite{hastie} 
(implemented using scikit-learn~\cite{scikitlearn}).
Results from the three regression methods are compared.
Standard ML training procedures were used in this study, as described in Ref.~\citen{hastie,tanaka2015}.
We divided our data set into a training set and a test set.
\footnote{We made a 70/30 training/test split. We use the training set to perform cross-validation.}
We trained our ML model and tuned the model's hyperparameters using k-fold cross-validation using the training data. 
We reported the model performance using the test set.
Since our test set is hidden from the training of the ML model, the test set serves as unseen experimental data.
The calculated test set accuracy approximates the expected prediction accuracy of the outcomes of new reactions or experiments.
See Supplementary Information for details.

We focus on reactions that form one product that maintains an intact carbon chain and omit those reactions where decomposition to multiple products with extended carbon chains occurs. This was done to focus on accurately predicting selective reactions on bare and oxygen covered Au.

ML models map the molecular representation of the reactants to the chemical space values of the products, ${{\bm{\mu}}}$$_{prod}$. 
For the case of kernel ridge regression,
\begin{equation}
 \hat{\bm{\mu}}_{prod}
(\bm{\kappa}) = 
 \sum^{N}_{i=1} 
 \bm{\alpha}_i
 \exp\left(
 -\frac{1}{2\sigma^2}{|\bm{\kappa}-\bm{\kappa}_{i}|}^2
 \right)
\label{eqn1}
\end{equation}

where the coefficients $\bm{\alpha}_i$ and parameter $\sigma$ are determined using kernel ridge regression with k-fold cross-validation~\cite{hastie, rupp2012}.
$\hat{\bm{\mu}}_{prod}$ is the ML prediction of  $\bm{\mu}$ for the product.

After predicting the products' chemical space values, 
the molecular structure must be reconstituted. 
Because there exists an enormous number of molecular structures and our predictions have uncertainty,
any given chemical space value prediction could plausibly correspond to dozens or even hundreds of compounds
in the same region of chemical space. 
Moreover, calculating the chemical space value of all the known molecules in chemical databases 
in order to find a match
%and creating a dictionary to map the chemical space value to the molecular structure 
is computationally prohibitive. 
% In addition, our machine learning predictions have some uncertainty.
%Therefore, errors will arise when matching the chemical space values of the molecules
%in a database.
Therefore we constrained the number of possible reaction outcomes 
by exploiting known organic reaction mechanisms to identify possible products associated with a given reaction. 
In addition, we imposed a heavy atom count penalty on the set of possible
products for a given reaction.
This penalizes predictions that do not conserve the number of heavy atoms for a given reaction.
Ultimately, we use a combination of rule-based approaches~\cite{Jianfeng2018} (a `reaction tree') and ML tools (see Fig.~\ref{fig1}). 

%While the machine learning model provides accurate chemical space predictions, chemical space value is a single variable that cannot uniquely specify the enormous variety of possible structures and compositions. This one-to-many relationship means that many compounds can have very similar chemical space values. Uncertainty in our predictions then means that many compounds can plausibly correspond to dozens or even hundreds of possible molecules.

%However, possible reaction products are constrained by stoichiometry and the possible reaction steps that Au(111) and Au(110) surfaces can catalyze. Our SMARTS patterns follow these limitations so the reaction trees restrict machine learning predictions to plausible molecules. Only a small subset of all possible molecules can be reached given these restrictions so reaction trees reduce ambiguity when converting one-dimensional chemical space values to a specific molecule and therefore improve prediction accuracy.

%\begin{itemize}
%    \item Robert's explanation of reaction trees %[appropriate for methods section]
%    \item database used
%    \item chemical rules used 
%\end{itemize}

%%%%%%%%%%%%%%%%%%%%%%%%%%%%% RAH %%%%%%%%%%%%%%%%%%%%%%%
We construct reaction trees by applying SMARTS patterns, each encoding a molecular transformation modeled after a known organic reaction mechanism. 
Patterns are either a \textit{monomer} reaction, transforming a single molecule, or a \textit{coupling} reaction, transforming two molecules. Reaction trees are then grown in three stages: 1) apply all monomer patterns to the reactants to obtain primary products, then to those products to obtain secondary products, and so on; 2) apply all coupling patterns to each possible pair of reactions from step 1; 3) apply all monomer patterns to the coupling products as described in step 1. 
This three-stage process generates a tree of products, each resulting from a sequence of reaction mechanisms. 
The specific SMARTS patterns are included in the Supplementary Information. 
Each pattern is tuned to avoid highly unstable products due to ring strain and other effects, and to respect stoichiometry. 
We then select the compound in this tree whose chemical space value best matches the machine learning prediction. 
Reaction trees greatly improve accuracy by restricting predictions to compounds whose structures and atomic compositions are consistent with the reactants, which is a small subset of all known compounds.
%For example, there are \todo{large number} of possible compounds with up to  \todo{small number} non-hydrogen atoms,\todo{reference} but typical reaction trees only contain 100 to 1,000 compounds.

\begin{figure*}[ht]
  \includegraphics[width=0.99\textwidth, keepaspectratio]{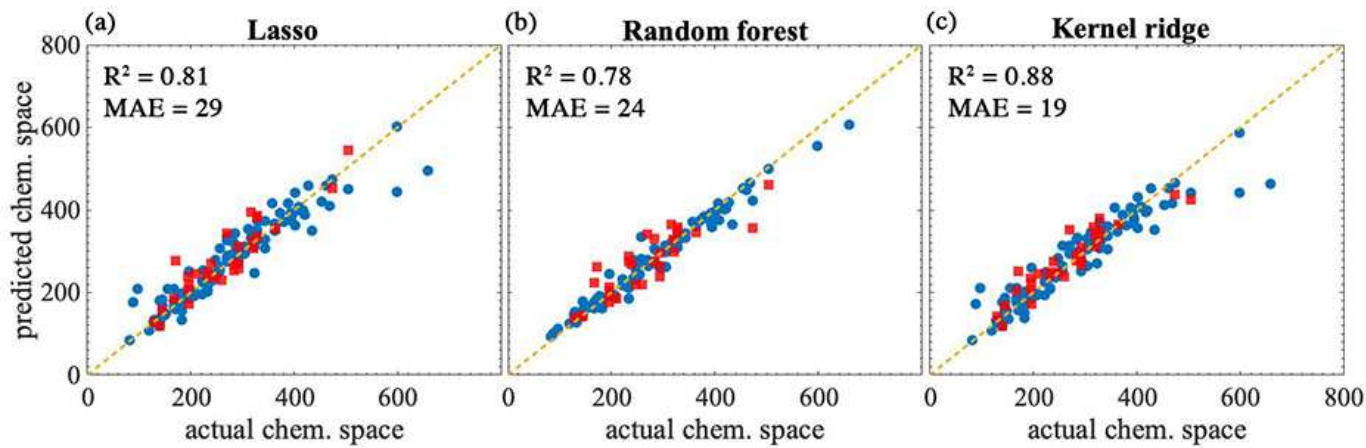}
  \caption{ \label{fig4} Machine learning predictions of the norm of the chemical space values for (a) lasso (b) random forest and (c) kernel ridge regression. 
  The horizontal axis is the norm of the actual chemical space value of the product of a reaction. 
  The vertical axis is the machine learning prediction of the norm of the chemical space value.
  Training (test) data are labelled with blue circles (red squares). 
  The coefficient of determination (R$^2$) and the mean absolute error (MAE) are shown in the panels.
}
\end{figure*}

To improve accuracy we apply two additional restrictions to reduce the number of compounds in each tree. 
First, trees are grown to include products that can be reached with a maximum of four reaction steps from the products (there are a few exceptions where trees are grown to 8 reaction steps, see Supplementary Information for details). 
%[note: add discussion on exception with tree depth 8]
This restriction mitigates the exponential growth of trees with the maximum number of allowed reactions. 
Second, some SMARTS patterns are only applied if overall `self-consistency' criteria are met to prevent the formation of products with unfeasible structures. 
For example, activation energies are significantly lower for alcohol dehydrogenation than abstracting two hydrogen atoms from a C--C bond, so we only allow C=C bond formation from C--C bonds for intermediates lacking aliphatic alcohol groups. 
%In addition to the local environments specified by the SMARTS patterns, these two additional restrictions decrease tree sizes.
%
Furthermore, not all of the generated SMILES strings exist in chemical databases. Consequently, the list of generated strings is filtered by those molecules that exist
in the PubChem database and therefore likely to be stable.

\section{Results}

\subsection{Chemical space}
We demonstrate that the chemical space values of different molecules, shown in Fig.~\ref{fig2}(a), encode the chemical behavior of these molecules.
We consider a set of molecules of distinct types and varying number of carbon atoms. 
We generate chemical space values
\footnote{Here we use the chemical space value based on the BoB representation modified using the bond energy.} 
for these structures and project the resulting vectors onto two dimensions using t-SNE~\cite{tsne}. 
Our chemical space representation is able to differentiate functionality in molecules as demonstrated by the clustering of molecules of varying carbon chain length and distinct functional groups.
%along a unique 2-d vector 
%and increasing . 
We consider linear alcohols, branched alcohols, terminal aldehydes and alkyl esters (shown in Fig.~\ref{fig2}(a)).
Our chemical space representation accounts for changes in the length of the carbon chain within a molecular class. This is demonstrated by the systematic spacing of points for a given class of molecules. 

%This suggests that
%the chemical space representation encodes chemical information.
%Figure~\ref{fig2}(a) shows several types of molecules, with an increasing number of carbon atoms,  and their corresponding chemical similarity. 
%This illustrates that molecules which are close in chemical space are also chemically similar.

In addition, we define the difference between chemical space vectors of products and reactants, $\bm{\delta}_{rxn}=\bm{p}-\bm{r}$
\footnote{Here we used both the chemical space value based on the BoB representation modified with the oxidation state and the original BoB representation. They both yield similar results.}.
For our set of reactions (with varying reaction types) we project $\bm{\delta}_{rxn}$ onto two dimensions using t-SNE (see Fig.~\ref{fig2}(b)).
t-SNE is often used as a dimensionality reduction tool and is only used here to visualize our high-dimensional chemical space vector representation of molecules as well as reactions. 
Our chemical space representation can effectively account for different classes of chemical transformation as evidenced by the clustering of reaction types. 
For reaction types 
that only have one possible transformation of reactant functionality there is only one cluster of reactions, 
for example, deoxygenation and molecular desorption. 
For reaction types 
%in our dataset 
where there are several possible transformations of reactant functionality, there are several clusters of reactions, for example
partial oxidation (see Supplementary Information for additional reaction types). 
This clustering suggests that our chemical space representation can account for the class of chemical reaction and the transformation of reactant functionality. 
We exploit this relationship to predict reaction outcomes.

\subsection{Machine learning predictions}
%
% We use the training set to train a machine learning model (using k-fold cross validation) 
% to predict the products of catalytic reactions.
%
ML predictions of the chemical space values of the reactions outcomes are shown in Fig.~\ref{fig4}. 
% We focus on reactions that form one product that maintains an intact carbon chain and omit those reactions where decomposition to multiple products with extended carbon chains occurs. This was done to focus on accurately predicting selective reactions on bare and oxygen covered Au.
%
%\todo{I wonder if this would be better in the methods %somewhere.---> YES!}
%
Kernel ridge regression performs  better than random forest regression,
while lasso has the worst performance.
Random forest regression suggests that the top three descriptors are the BoB descriptors based on oxidation number, bond energy and atomic number for  Reactant 1
(see Supplementary Information for details).
Despite the small data set size, all models achieve good prediction accuracy for both training and test data.

For a given reaction, 
we can select the predicted product from the reaction tree by calculating the following:
${{\Lambda} = |\hat{\bm{\mu}}_{prod} - \bm{\mu}_{tree}|}$,
where $\bm{\mu}_{tree}$ is a list of chemical space values of molecules in a reaction tree. 
We then determine ${\textrm{argmin}_{t \in tree}}\left| \hat{\bm{\mu}}_\textrm{prod} - \bm{\mu}_t \right|$, where $\bm{\mu}_t$ is the chemical space value of molecule $t$ in the tree,
that is, we select the reaction tree products that are the most chemically similar to the ML prediction.

We can choose more than one molecule to improve the chances of a correct match.
This accounts for the uncertainty of ML prediction.
Increasing the number of guesses $\gamma$ from one to three changes the prediction accuracy from 85\% to 90\% for all the reactions in the data set. 
Therefore, for a particular application we can choose a value of $\gamma$ that allows the proper trade-off between having a high certainty that the actual product is one of the guesses and having a low number of possible products to consider. 
We define prediction accuracy as the fraction of cases where at least one of the guesses is a correct prediction.
% ADDED my TR April 25:
Above we reported the prediction accuracy of the entire data set. 
If we examine the test set only, we can find up to 100$\%$ prediction accuracy for the test set. 
However, this strongly depends on how we make the training/test set split. 
We believe that our trained model is not overfitting and can generalize reasonably well. 
Therefore, for the remainder of the discussion we report the prediction accuracy on the entire data set.
%
%\todo{Seems like this should go either before or after the whole %$\theta$ discussion.}

We can improve prediction accuracy by adding constraints to
$\Lambda$. 
We impose a penalty on the difference in the number of each heavy atom (C, N, O, etc.) before and after the reaction,
that is, for a given reaction we penalize those reaction tree products   that do not conserve the number of heavy atoms. 
We define a vector $\textbf{H}$ describing the number of heavy atoms of reactants or products.
$\Delta\textbf{H}$ is the difference in the number of heavy atoms of the reactants and the products. 
We build a ML model to predict $\Delta \textbf{H}$, thereby estimating
the number of heavy atoms of the products for a given reaction.
We tune the heavy atom number penalty $\Delta\hat{\textbf{H}}$ by incorporating a hyper-parameter $\theta$ that adjusts the relative importance of
$\Delta\hat{\textbf{H}}$ with respect to the chemical space difference in the expression below:
\begin{equation}
 {\Lambda}_{t}(\theta) =  |\hat{\bm{\mu}}_{prod} - \bm{\mu}_{t}| + \theta \sum{|\Delta\hat{\textbf{H}}_t|}
\label{costfunc}
\end{equation}
\begin{equation}
  \textrm{argmin}_{t \in tree}[{{\Lambda(\theta)}}]
\label{costfunc2}
\end{equation}
%where $\vec{P}_{prod}$
%is the machine learning prediction of the chemical space value and
%where $\vec{P}_{tree}$ is the set of possible products given by the reaction tree.
%
We find chemical space similarities for $t$ molecules in the reaction tree and impose the heavy atom conservation rule.
We choose the argument of the $\Lambda$ list that returns the minimum value. 
We can also rank the molecules in increasing order of chemical similarity difference from the minimum.
%
%$\theta$ is a tuning parameter to be optimized and $\Delta\textbf{H}$ is the penalty on the heavy atom counts, a vector describing the difference in the number of heavy atoms e.g. C, N and O.
%These vectors are determined using machine learning.
%See the supplementary information for details.
%
%Prediction accuracy here is defined to be the fraction of times at least one of the guessed molecules yields a correct match.
%However, a larger $\gamma$ increases the number of ML predictions, giving rise to uncertainty in the precise outcome of a reaction. 
%Nevertheless, this ranking of  potential reaction outcomes will be a useful tool for experimentalists.
%In the limit of large $\gamma$ (approaching all the possible products in a reaction tree) we always find the correct outcome of the reaction, albeit from potentially large set of reactants.
%
%
\begin{figure}[ht]
\begin{center}
  \includegraphics[width=0.35\textwidth, keepaspectratio]{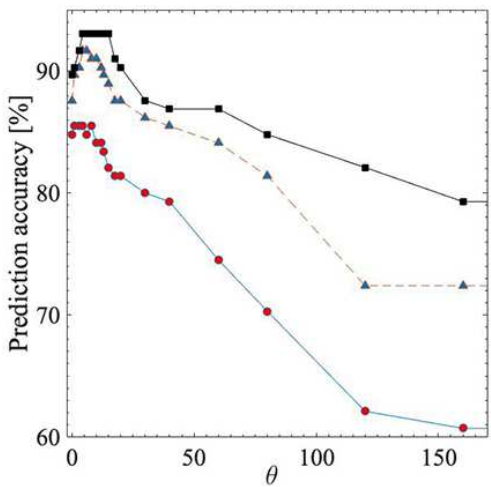}
  \caption{ \label{fig5} 
  ML prediction accuracy versus the heavy atom count penalty for number of guesses, $\gamma$ = 1 (circles), 2 (triangles) and 3 (squares).
  The heavy atom count penalty is tuned with parameter $\theta$.
  }
  \end{center}
\end{figure}

Fig.~\ref{fig5} illustrates how the prediction accuracy varies as $\theta$ is tuned. 
When $\theta$ is close to zero, the heavy atom penalty is neglected. 
For large $\theta$, the heavy atom penalty dominates the $\Lambda$ function. 
We choose a $\theta$ value that optimizes the prediction accuracy.
Imposing the heavy atom penalty with the optimal weighting, $\theta = 4$, increases the prediction accuracy to 93\% for $\gamma$ = 3.
Since the heavy atom counts for the products are predicted using ML,
more than one ML prediction is leveraged to 
improve prediction accuracy. 
This is analogous to the performance enhancement seen with ensemble methods~\cite{hastie2001}.
%
%\todo{It feels more like an ensemble method than boosting to me, but if you like the analogy to boosting that's fine.-->TR: Okay it's not sequential like boosting so the more general term 'ensemble' is better}
%
%\todo{Should we have a separate discussion header here? Seems like it's already divided into results and discussion. --> TR: I've seen this done before in this journal. we can do it}
%CRO comments:
%
\section{Discussion}
We find an overall high prediction accuracy of 86\%, 91\% and 93\% for $\gamma$ = 1, $\gamma$ = 2, and $\gamma$ = 3, respectively. Some of the error in prediction accuracy is due to the fact that a tree depth of 4 is not deep enough for the experimentally observed product to appear. We tested a larger tree depth of 8 for reactions where the observed product did not appear in the trees of depth 4, and this resulted in correct predictions from our methodology in all of these cases (for $\gamma$ = 2 and 3). 
% This is true for gamma=3 but there's an error for
% gamma=2,3 with one error for gamma=1
With this larger depth for these specific cases, the prediction accuracy increased to 89\%, 95\% and 97\% for $\gamma$ = 1, $\gamma$ = 2, and $\gamma$ = 3, respectively. Generally, in the limit of very small tree depths the observed product will not appear. Similarly, in the limit of very large tree depths the density of molecules in chemical space may become too large to properly identify the observed product. This suggests that properly tuning the reaction tree depth can play a key role in the prediction accuracy when predicting reactions where no experimental confirmation exists yet. 
We have found that a tree depth of 4 is sufficient to give reasonably high accuracy and consider this case for the rest of this work.

Our ML framework yields better predictions for some types of reactions or molecules.
This may be due to our choice of chemical space descriptors. 
For instance, the
predictions have a high accuracy for all surface reactions tested on oxygen covered Au and bare Au surfaces with the exception of cases with carbon rings when an extended tree depth is used.
%
% {TR: is this needed??}
%when an extended tree depth is used. 
%
The prediction accuracy for carbon rings was 76\% for one guess, as compared to 95\% for other structures. The high error for carbon ring structures can be improved by using two or three guesses, which brings the accuracy up to 89\% and 96\%, respectively. This suggests that our chemical descriptors can locate the region of chemical space in which carbon ring structures exist but cannot accurately distinguish between the very high density of isomers of carbon ring structures in that space. In order to address the error in ring structure predictions, new descriptors would need to be developed to effectively account for local ring structure. The errors for structures without a carbon ring occur when the reaction involves a relatively large number of elementary steps. The error in predicting complex surface reactions would likely be reduced by providing additional data that involve many elementary surface reaction steps. 
Nevertheless, the predictions are generally accurate but currently have higher error rates for complex ring structures and complex, multi-step surface reactions. 

We were able to predict a variety of reaction classes with reasonable certainty on both clean and oxygen covered Au. Our approach was effectively able to ``learn'' complex oxidation chemistry from a limited data set due to (i) a representation of molecules in chemical space and (ii) general reaction rules that apply to all surfaces, which constrain the chemical space for possible products. This provides a remarkable tool that may reduce the need for time consuming experimental measurements while properly accounting for known chemical pathways. We can accurately predict a wide range of reactions including: (i) desorption of oxygenates on Au, (ii) deoxygenation of oxiranes on Au, (iii) desulfurization of thiols on Au, (iv) oxidative coupling of oxygenates on O/Au, (v) partial oxidation of oxygenates on O/Au and (vi) oxidative dehydrogenation of oxygenates on O/Au. Overall, we developed a methodology that allows for the effective use of a ML approach to complement chemical insight for determining the outcome of catalytic reactions.

We emphasize that we were able to predict both desorption processes and reactions with reasonable certainty on clean Au. 
%\todo{We should lay out a specific example somewhere.}
Molecular desorption typically occurs for organics on bare Au, but our ML model is able to accurately predict reactions for the following unique cases: (i) the deoxygenation of the oxiranes 2-methyloxirane, 2-phenyl oxirane, and oxirane; and (ii) the desulfurization of thiols: ethanethiol, benzenethiol, 2-methylpropane-2-thiol.
The ML model ``learned'' oxygen chemistry and was able to transfer that knowledge to chemically similar reactions involving sulfur molecules. 
That is, our ML approach demonstrated its own chemical insight that proved useful for generating successful predictions of catalytic reaction outcomes.
The success of this machine learning approach may lie in the ability of these methods to identify patterns and relationships in a high-dimensional space of chemical properties, which is difficult for humans to directly perceive.

We anticipate that this approach can be applied to other monometallic surfaces, such as Ag or Pt. However, the overall success of ML models will depend heavily on whether sufficient data exist. Given a enough data for several surfaces, it may be possible to develop a single, general model that encompasses multiple surfaces, which would decrease the amount of data needed for a single surface. 
This would likely require the development of descriptors that capture the appropriate chemical properties of distinct surfaces.

\section{Conclusion}
We used ML tools to predict the outcomes of small molecule reactions on the surface of bare gold and oxygen-covered gold.
%\todo{We looked at other cases besides O-covered, right? TR: YES!}
We showed that ML can be a useful guide in predicting the outcome of catalytic reactions for small data sets with high accuracy.
Our model relies on a chemical space that is suitable for predicting chemical reaction outcomes. 
We combined ML tools with traditional reaction rules,
which we used to transform chemical space predictions of products to the corresponding molecular structures.
%We found introducing a heavy atom penalty could
%improve the machine learning prediction accuracy of the reaction outcomes.
This work sets the stage for future catalysis studies of a variety of metals and coverage by species other than oxygen. 
Furthermore, our ML study provides a framework for inverse design, where we can perform a high-throughput search for a desirable reactant (or reactants) with which to create a particular product. 
This work shows particular promise since millions of training examples are not needed as is often the case in ML methods.
%In addition, this study sets the stage for a data-driven approach to an investigation of reaction pathways of  catalytic reactions.
In particular, our study provides a useful tool that could guide catalysis experiments, either complementing or supplementing traditional chemical intuition.
With proper modification this approach could be useful in the development of industrially relevant catalytic reactions.
%
%\todo[inline]{We need some more big-picture stuff here (what's new and exciting, what are some implications going forward), instead of just a summary. This might be worth discussing. }

\section*{Conflicts of interest}
There are no conflicts of interest to declare.

\section*{Acknowledgements}
The authors thank Pavlos Protopapas, Koji Tsuda, Vinothan Manoharan, Jonathan Rhone, Marios Mattheakis, Daniel Larson and Jovana Andrejevic for helpful discussions.
T.D.R. is supported by the Harvard Future Faculty Leaders Postdoctoral Fellowship.
%We acknowledge support from ARO MURI Award W911NF-14-0247. 
R.H., C.R.O. and M.M.M were supported as part of Integrated Mesoscale Architectures for Sustainable Catalysis (IMASC), an Energy Frontier Research Center funded by the U.S. Department of Energy, Office of Science, Office of Basic Energy Sciences under Award Number DE- SC0012573.

%\bibliography{MMM.bib,references.bib,Bibtex_chris.bib}
%\bibliography{references.bib,MMM.bib,Bibtex_chris.bib}

\providecommand*{\mcitethebibliography}{\thebibliography}
\csname @ifundefined\endcsname{endmcitethebibliography}
{\let\endmcitethebibliography\endthebibliography}{}

\bibliographystyle{rsc} %the RSC's .bst file

\end{document}